\documentclass{naturemod}
\usepackage{graphicx}
\usepackage{pifont}
\usepackage{dcolumn}
\usepackage{bm}

\usepackage{amssymb}
\usepackage{amsmath}
\usepackage{mathtools}
\usepackage{color}
\usepackage{array}

\title{Real-space imaging of non-collinear antiferromagnetic \mbox{order} with a single spin magnetometer}

\author{I. Gross$^{1,2,}$\footnote{These authors contributed equally to this work}, W. Akhtar$^{1}$\footnotemark[\value{footnote}], V. Garcia$^{3}$, L.~J.~Mart\'{\i}nez$^{1}$, S. Chouaieb$^{1}$, K.~Garcia$^{3}$, C.~Carr\'et\'ero$^{3}$, A.~Barth\'el\'emy$^{3}$, P.~Appel$^{4}$, P.~Maletinsky$^{4}$, J.-V. Kim$^{5}$, J. Y. Chauleau$^{6,7}$, N. Jaouen$^{7}$, M. Viret$^{6}$, M.~Bibes$^{3}$, S. Fusil$^{3}$ and V.~Jacques$^{1}$}

\begin{document}
\maketitle
\begin{affiliations}
\item Laboratoire Charles Coulomb, Universit\'e de Montpellier and CNRS, 34095 Montpellier, France
\item Laboratoire Aim\'{e} Cotton, CNRS, Universit\'{e} Paris-Sud, ENS Cachan, Universit\'{e} Paris-Saclay, 91405 Orsay, France
\item Unit\'e Mixte de Physique, CNRS, Thales, Univ. Paris-Sud, Universit\'e Paris-Saclay, 91767 Palaiseau, France
\item Department of Physics, University of Basel, Klingelbergstrasse 82, Basel CH-4056, Switzerland
\item Centre de Nanosciences et de Nanotechnologies, CNRS, Universit\'e Paris-Sud, Universit\'{e} Paris-Saclay, 91405 Orsay, France
\item SPEC, CEA, CNRS, Universit\'e Paris-Saclay, 91191 Gif sur Yvette, France
\item Synchrotron SOLEIL, 91192 Gif-sur-Yvette, France
\end{affiliations}
\newpage
{\bf While ferromagnets are at the heart of daily life applications, their large magnetization and resulting energy cost for switching bring into question their suitability for reliable low-power spintronic devices. Non-collinear antiferromagnetic systems do not suffer from this problem and often possess remarkable extra functionalities: non-collinear spin order~\cite{Coey} may break space-inversion symmetry~\cite{Kimura,Cheong} and thus allow electric-field control of magnetism~\cite{Lottermoser,Heron}, or produce emergent spin-orbit effects~\cite{Nayak} which enable efficient spin-charge interconversion~\cite{Zhang}. To harness these unique traits for next-generation spintronics, the nanoscale control and imaging capabilities that are now routine for ferromagnets must be developed for antiferromagnetic systems. Here, using a non-invasive scanning nanomagnetometer based on a single nitrogen-vacancy (NV) defect in diamond~\cite{Maze,Gopi,Rondin_2014}, we demonstrate the first real-space visualization of non-collinear antiferromagnetic order in a magnetic thin film, at room temperature. We image the spin cycloid of a multiferroic BiFeO$_3$ thin film and extract a period of $\sim 70$~nm, consistent with values determined by macroscopic diffraction~\cite{Sosnowska,Gukasov2008}. In addition, we take advantage of the magnetoelectric coupling present in BiFeO$_3$ to manipulate the cycloid propagation direction by an electric field. Besides highlighting the unique potential of NV magnetometry for imaging complex antiferromagnetic orders at the nanoscale, these results demonstrate how BiFeO$_3$ can be used as a versatile platform for the design of reconfigurable nanoscale spin textures.}

Nearly $90\%$ of known magnetic materials have dominant antiferromagnetic interactions, resulting in no or very small magnetization, and most are also insulators~\cite{Coey}. This strongly impedes their investigation, especially when the magnetic order needs to be mapped at the nanoscale. While magnetic force microscopy~\cite{Hartmann1999} or X-ray photoemission electron microscopy~\cite{XPEEM1} can reach a spatial resolution of a few tens of nm, their sensitivities are not compatible with the detection of weak magnetic signals commonly involved in antiferromagnets. Spin-polarized scanning tunnelling microscopy can resolve the magnetic moments of single atoms~\cite{Wiesendanger} but is only applicable to conductive systems. Therefore, the spin texture of the vast majority of magnetically ordered materials cannot be directly imaged at the nanoscale. This is increasingly problematic since materials with complex antiferromagnetic orders show very appealing functionalities, which are absent in ferromagnets, and start to be exploited in a new generation of low-power spintronic devices~\cite{AFreview}. 

Typical examples are multiferroics, in which antiferromagnetism coexists with ferroelectricity, enabling an efficient electrical control of magnetization through magnetoelectric coupling~\cite{Cheong,Lottermoser,Heron}. Bismuth ferrite BiFeO$_3$ (BFO) is such a multiferroic material~\cite{Catalan}, which is currently emerging as a unique platform for spintronic~\cite{Heron} and magnonic devices~\cite{Rovillain_2010} because its multiferroic phase is preserved well above room temperature. However, while the ferroelectric properties of BFO have been widely investigated by piezoresponse force microscopy (PFM), revealing unique domain structures and domain wall functionalities\cite{Balke2012,Catalan2012}, the corresponding nanoscale magnetic textures and their potential for spin-based technology still remain concealed. In this work we demonstrate the first real-space imaging and electric field manipulation of complex antiferromagnetic order in a BFO thin film by using an atomic-sized magnetometer based on a single NV defect in diamond.

\begin{figure}
		\centering
		\includegraphics[scale=0.5]{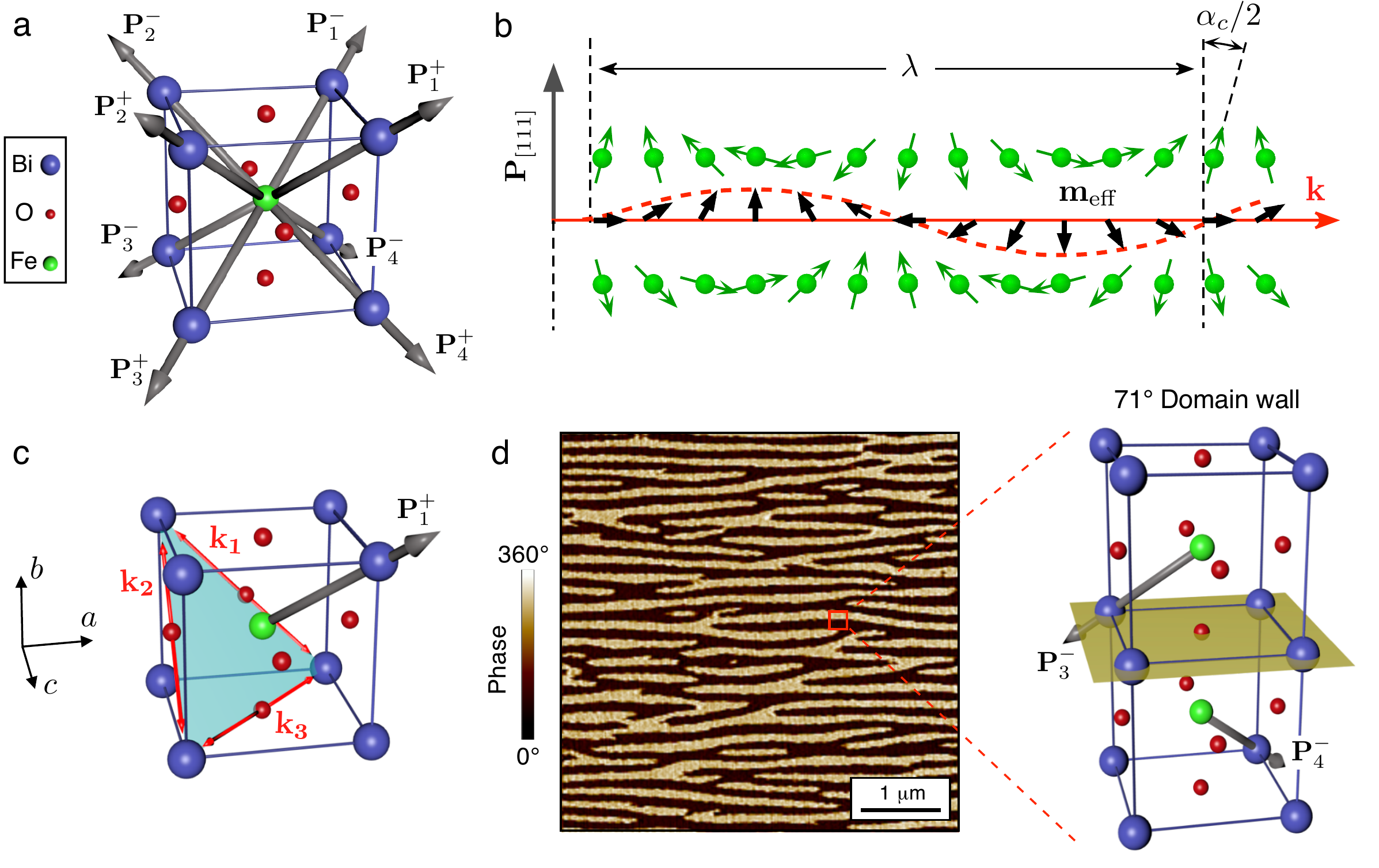}
		\caption{{\bf Ferroelectric and magnetic order in BiFeO$_3$.} {\bf a,} Pseudocubic unit cell of BiFeO$_3$ showing the possible variants of the ferroelectric polarization $\mathbf{P}_i^{\pm}$ pointing along the eight $[111]$ directions. {\bf b,} Schematic representation of the spin cycloid. Magnetoelectric coupling induces a cycloidal rotation of Fe$^{3+}$  spins (green arrows). The canted antiferromagnetic alignment between consecutive atomic layers, characterized by the angle $\alpha_c$, results in an effective magnetic moment $\mathbf{m}_{\rm eff}$ describing a cycloid with a wavelength $\lambda$ (black arrows). The propagation direction of the spin cycloid $\mathbf{k}$ is normal to the ferroelectric polarization vector $\mathbf{P}$. {\bf c,} Representation of a given variant of the ferroelectric polarization ($\mathbf{P}_1^{+}$ along the [111] axis) together with the three possible propagation directions of the spin cycloid \textbf{k$_1$} $\parallel$ [$\bar{1}1$0], \textbf{k$_2$} $\parallel$ [$01\bar{1}$]  and \textbf{k$_3$} $\parallel$ [$10\bar{1}$].  {\bf d,} Striped pattern of ferroelectric domains in the (001)-oriented BiFeO$_3$ thin film probed by piezoresponse force microscopy (PFM). The right panel sketches the two pristine variants of ferroelectric domains ($\mathbf{P}_3^{-}$ and $\mathbf{P}_4^{-}$) separated by $71^{\circ}$ domain walls. The sketches in {\bf a}, {\bf c} and {\bf d} are in top view with a small tilt.}
	\end{figure}
	
Bulk BFO crystallizes in a slightly-distorted rhombohedral structure, but is commonly described by the pseudocubic unit cell shown in Fig.~1a. The displacement of Bi ions relative to the FeO$_6$ octahedra gives rise to a strong ferroelectric polarization ($100 \ \mu$C/cm$^2$) along one of the $[111]$ directions~\cite{Catalan}. This system is complex as the eight possible polarization orientations $\mathbf{P}_i^\pm$ give rise to three types of ferroelectric domain walls ($71^{\circ}$, $109^{\circ}$, or $180^{\circ}$). From the magnetic point of view, BFO was initially thought to be a conventional G-type antiferromagnet~\cite{Park} but high-resolution neutron diffraction later revealed a cycloidal antiferromagnetic order~\cite{Sosnowska,Gukasov2008} with a characteristic wavelength of $\lambda \sim 64$~nm [Fig.1b]. The spin cycloid propagation direction and the ferroelectric polarization vector are normal to each other and are linked by magnetoelectric coupling. In addition, the rhombohedral symmetry of BFO allows three equivalent propagation directions of the cycloid ($\mathbf{k}_1,\mathbf{k}_2,\mathbf{k}_3$) for a given variant of ferroelectric domain~\cite{Gukasov2008,Park} [Fig. 1c]. 
	
A $32$-nm-thick BFO(001) film was grown by pulsed laser deposition on a DyScO$_3$(110) orthorhombic substrate, using an ultrathin buffer electrode of SrRuO$_3$ [cf. Methods and Extended Data Fig.~1]. Epitaxial strain leads to an array of striped ferroelectric domains whose typical width is in the range of $\sim100$~nm [Fig.~1d]. In-depth PFM and X-ray diffraction analysis reveal that only two variants of polarization coexist [$\mathbf{P}_3^-$ and $\mathbf{P}_4^-$ in Fig.~1d], separated by $71^{\circ}$ domain walls [cf. Methods and Extended Data Fig.~2 and 3]. In thin films the spin cycloid can be modulated or even destroyed by epitaxial strain. Considering the low lattice mismatch between BFO and DSO ($\sim 0.4\%$), the cycloidal antiferromagnetic order is however expected to be preserved in the studied epitaxial thin film~\cite{Sando2013}. 
	
The spin texture of the BFO sample was investigated through stray field measurements using a scanning nanomagnetometer based on a {\it single} NV defect in diamond~\cite{Maze,Gopi,Rondin_2014}. This point-like impurity can be exploited for quantitative magnetic field imaging at the nanoscale by recording Zeeman shifts of its electronic spin sublevels through optical detection of the electron spin resonance (ESR). For the present study, a single NV defect placed at the apex of a nanopillar in a diamond scanning-probe is integrated into an atomic force microscope, which allows scanning the NV defect in close proximity to a sample\cite{Maletinsky_2012} [Fig.~2a]. At each point of the scan, optical illumination combined with radiofrequency (RF) excitation enable measuring the ESR spectrum of the NV defect by recording its spin-dependent photoluminescence (PL) intensity [Fig.~2b]. Any magnetic field emanating from the sample is then detected through a Zeeman shift of the ESR frequency, which is simply given by $\Delta_z=\gamma_e B_{\rm NV}/2\pi$, where $\gamma_e/2\pi=28$~GHz/T is the electronic spin gyromagnetic ratio and $B_{\rm NV}$ is the magnetic field projection along the NV defect quantization axis. The resulting magnetic sensitivity is in the range of a few $\mu$T/$\sqrt{\rm Hz}$, while the spatial resolution is fixed by the distance $d$ between the sample and the NV spin sensor~\cite{Rondin_2014}. This key parameter is independently measured through a calibration process above the edges of an uniformly magnetized ferromagnetic wire~\cite{Tetienne2015}, leading to $d=49.0\pm 2.4$~nm [cf. Methods and Extended Data Fig.~4]. In the following, all experiments are performed under ambient conditions with a bias field $B_{\rm b}=1.4$~mT applied along the NV defect axis in order to determine the sign of the measured magnetic fields~\cite{Rondin_2014}. Such a bias field is weak enough not to modify the magnetic order in BFO.

\begin{figure}
		\centering
		\includegraphics[scale=0.5]{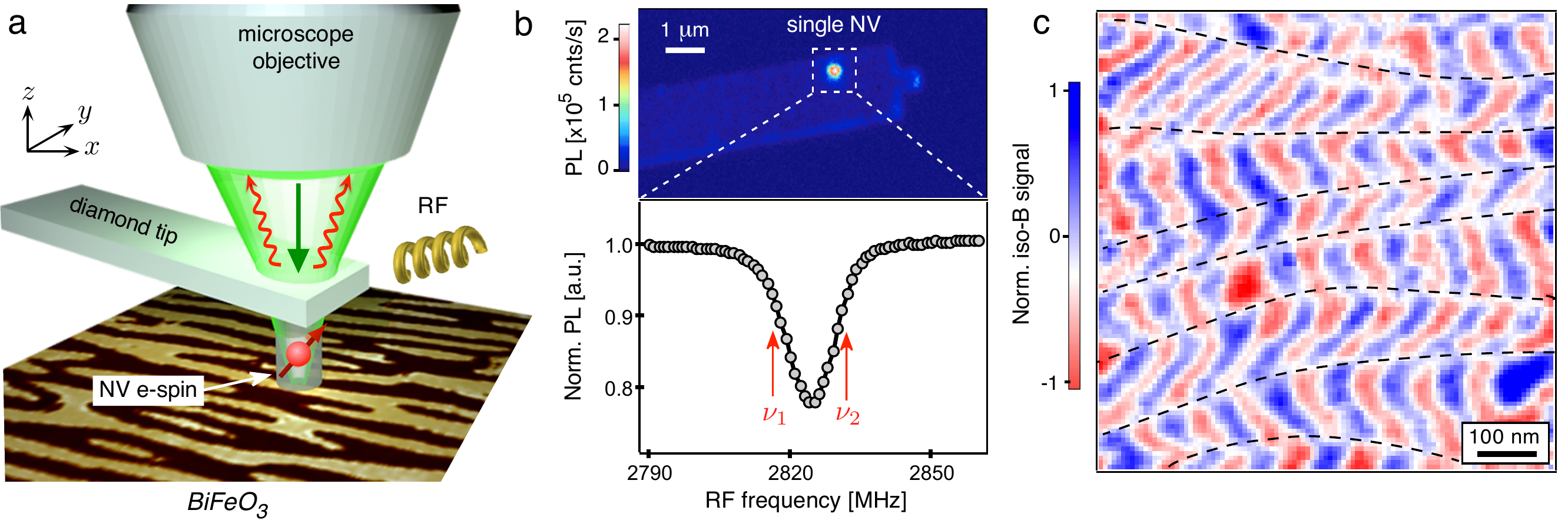}
		\caption{{\bf Mapping the magnetic texture of BiFeO$_3$ with NV magnetometry.} {\bf a,} The electronic spin of a single NV defect placed at the apex of a diamond scanning-probe tip is used as an atomic-sized magnetic field sensor. A microscope objective enables both to excite (green arrow) and collect the spin-dependent PL (red wavy arrows) of the NV defect, and a radiofrequency (RF) source is used to manipulate its electronic spin state [cf. Methods]. {\bf b,} (top panel) PL raster scan of the diamond scanning-probe showing the bright emission from a single NV defect. (bottom panel) Typical ESR spectrum recorded while applying a bias field $B_{\rm b}=1.4$~mT along the NV axis. The red arrows indicate the two RF frequencies $\nu_1$ and $\nu_2$ used for the {\it dual-iso-B} imaging mode. {\bf (c)} Magnetic field image recorded above the BFO film while operating the NV magnetometer in dual-iso-B imaging mode. The black dashed lines, which are drawn as a guide to the eye, are attributed to ferroelectric domain walls leading to abrupt rotations of the antiferromagnetic order.}
			\end{figure}
			
The scanning-NV magnetometer was first operated in the {\it dual-iso-B} imaging mode by monitoring the signal $\mathcal{S}=$PL($\nu_2)$-PL($\nu_1$), corresponding to the difference of PL intensity for two fixed RF frequencies, $\nu_1$ and $\nu_2$, applied consecutively at each point of the scan~\cite{Rondin_2014} [Fig.~2b]. A typical {\it dual-iso-B} image recorded above the (001)-oriented BFO thin film is shown in Fig.~2c. We observe a periodic variation of the magnetometer signal along the horizontal axis of Fig.~2c, which directly reveals the spatially oscillating magnetic field generated by the cycloidal modulation of the spin order. Moreover, the propagation direction of this spin cycloid is periodically modified along the vertical axis of Fig.~2c. The resulting zig-zag shaped magnetic field distribution mimics the shape and width ($\sim 100$~nm) of ferroelectric domains [Fig.~1d].

\begin{figure}
		\centering
		\includegraphics[scale=0.35]{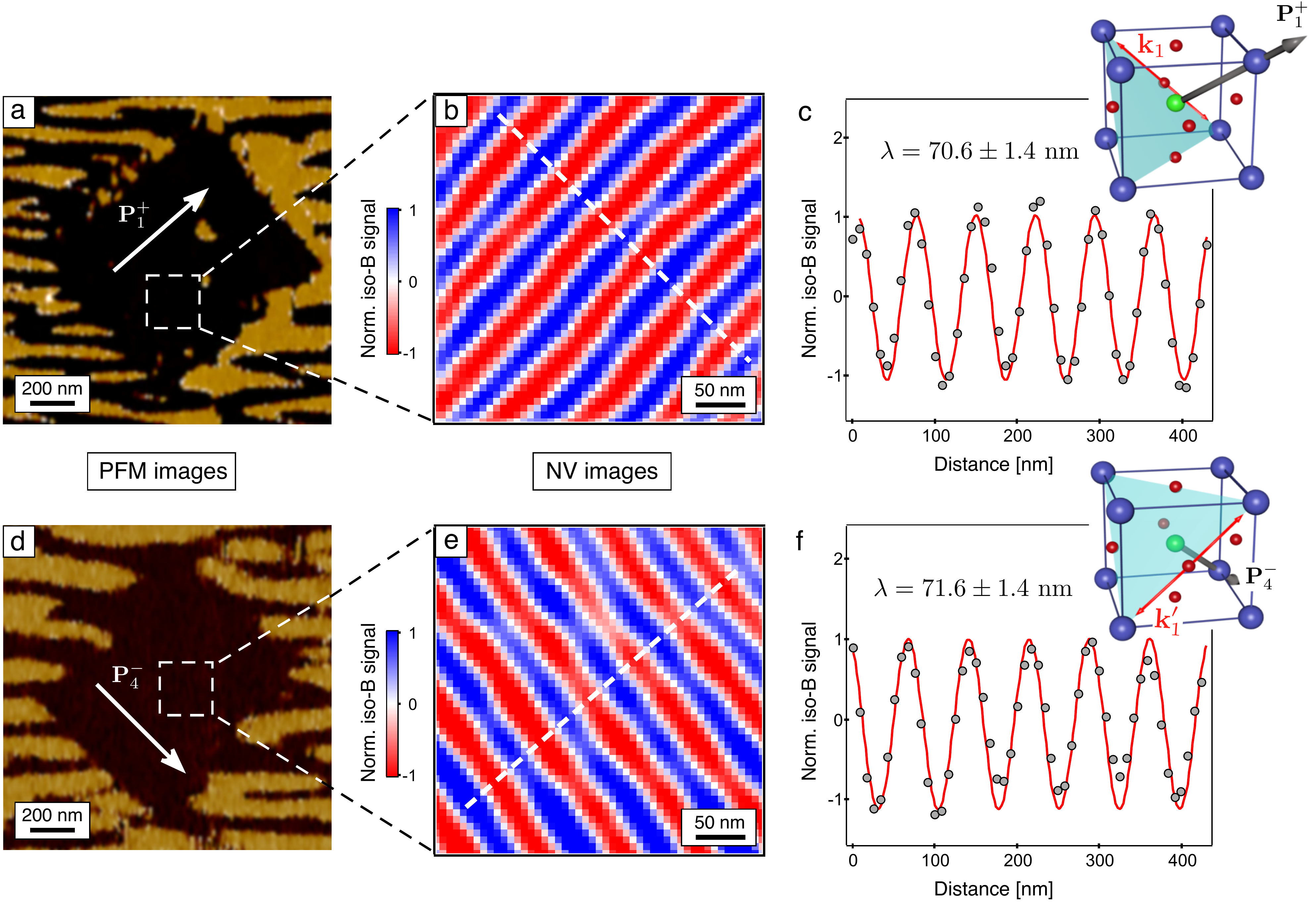}
		\caption{{\bf Electrical control of the spin cycloid. a, d,} In-plane PFM images of ferroelectric micron-sized domains with $\mathbf{P}_1^{+}$ and $\mathbf{P}_4^{-}$ polarizations, respectively. The white arrows indicate the in-plane projection of the ferroelectric polarization vector. {\bf b, e,} Corresponding magnetic field distributions recorded with the scanning-NV magnetometer operating in dual-iso-B imaging mode. {\bf c, f,} Linecuts of the magnetic field distribution along the cycloid propagation direction (white dashed lines in {\bf b} and {\bf e}, respectively). The cycloid wavelength $\lambda$ is extracted through a two-dimensional fit of the experimental data with a sinusoidal function (red solid lines). The standard error (s.e.) of the measurement ($\sim 2\%$) is limited by the calibration of the scanner. The insets show top view sketches of the ferroelectric polarization vector together with the propagation vector of the spin cycloid  $\mathbf{k}_1$ (in {\bf c})  and $\mathbf{k}_1^{\prime}$ (in {\bf f}).}
			\end{figure}
			
To gain further insights into the properties of the spin cycloid in this BFO thin film, PFM was used to design a single micron-sized ferroelectric domain from the as-grown striped pattern [Fig.~3a], taking advantage of the trailing electric field induced by the slow scan axis of the scanning probe [cf. Methods]. The magnetic field distribution recorded above such a ferroelectric monodomain exhibits a simple periodic structure, indicating the presence of a single spin cycloid [Fig.~3b]. Importantly, the (001) surface projection of the spin cycloid propagation direction is normal to that of the ferroelectric polarization vector $\mathbf{P}_1^+$. Among the three possible cycloid propagation directions, only $\mathbf{k}_1$ is normal to the (001) projection of $\mathbf{P}_1^+$, the other two lying at $45^{\circ}$ from the polarization vector [see inset in Fig.~3c]. We therefore conclude that the spin cycloid propagates along $\mathbf{k}_1$, {\it i.e.} in the plane of the BFO thin film. This result can be qualitatively explained by considering that epitaxial strain modifies the anisotropy along the film normal~\cite{Sando2013}. For BFO thin films grown on DSO, compressive strain induces an easy-plane contribution which stabilizes magnetic structures with their spins far from the $[001]$ direction. Thus, the three possible cycloidal directions see their degeneracy lifted and the one propagating along $[\bar{1}10]$ becomes energetically favorable~\cite{Sando2013}. Using a two-dimensional fit of the magnetic image with a sinusoidal function, we infer a characteristic wavelength $\lambda=70.6\pm1.4$~nm [Fig.~3c]. The slightly enhanced period compared to the bulk value ($\sim 64$~nm) is interpreted as due to the small compressive strain imposed by the substrate~\cite{Agbelele2017}. This result illustrates that the local magnetoelectric interaction between neighbouring atoms at the origin of the spin cycloid does not require thick films of BFO, {\it i.e.} with thicknesses well above its characteristic wavelength, as previously speculated~\cite{Bertinshaw2016}.

After demonstrating that the polarization and the cycloid propagation are intimately linked, we intend to manipulate electrically this cycloid propagation direction using the magnetoelectric coupling. To this end, we define another ferroelectric domain with an in-plane component of the polarization rotated by 90$^{\circ}$ [$\mathbf{P}_4^-$ in Fig. 1d]. The magnetic image shows that the propagation direction of the spin cycloid is also rotated by 90$^{\circ}$ with a very similar wavelength $\lambda=71.4$~$\pm1.4$~nm, once again corresponding to the propagation direction $\mathbf{k}_1^{\prime}$ lying in the (001) plane [Fig.~3e,f]. These experiments illustrate how magnetoelectric coupling can be used to efficiently control and manipulate the antiferromagnetic order in a BFO thin film. They also confirm that the abrupt rotations of the antiferromagnetic order observed in Fig.~2c are occurring at ferroelectric domain walls.

As a final experiment, a fully quantitative magnetic field image was recorded above the ferroelectic monodomain shown in Fig.~3a. Here the magnetic field component $B_{\rm NV}$ was obtained by measuring the Zeeman shift $\Delta_{z}$ of the NV defect electron spin sublevels at each pixel of the scan [cf. Methods]. The resulting magnetic field map indicates a modulation with a typical amplitude in the range of $\pm140 \ \mu$T [Fig.~4a]. In order to understand quantitatively such experimental data, we start by computing the stray field produced by the BFO sample. To this end, the spin cycloid is modeled by a rotating uncompensated magnetization vector $\mathbf{M}_{\rm eff}=\mathbf{m}_{\rm eff}/V$, where $V$ is the volume of the pseudo-cubic cell of BFO and 
\begin{equation}
\mathbf{m}_{\rm eff}(\mathbf{r^{\prime}}) = m_{\rm eff} \left[\cos(\mathbf{k_1}\cdot\mathbf{r^{\prime}})\mathbf{e_{k_1}}  +\sin(\mathbf{k_1}\cdot\mathbf{r^{\prime}})\mathbf{e_P}\right] \ .
\end{equation}
Here $||\mathbf{k_1}||=2\pi/\lambda$, $\mathbf{r^{\prime}}$ denotes the coordinate in the BFO sample, while $\mathbf{e_{k_1}}$ and $\mathbf{e_P}$ are orthogonal unit vectors oriented along the cycloid propagation direction $\mathbf{k_1}$ and the ferroelectric polarization $\mathbf{P}$, respectively [see Fig.~4b]. The uncompensated magnetic moment per Fe atom is given by $m_{\rm eff}=m_{\rm Fe}\sin(\alpha_c/2)$, where $m_{\rm Fe}=4.1 \ \mu_{\rm B}$ is the measured magnetic moment of Fe atoms in BFO at room temperature~\cite{Gukasov2008} and $\alpha_c$ is the canting angle between antiferromagnetically coupled Fe atoms [Fig.~1b]. This angle is directly deduced from the measured cycloid wavelength, leading to $\alpha_c=2^{\circ}$ and $m_{\rm eff}=0.07 \ \mu_{\rm B}$ [cf. Methods]. 

\begin{figure}
		\centering
		\includegraphics[scale=0.35]{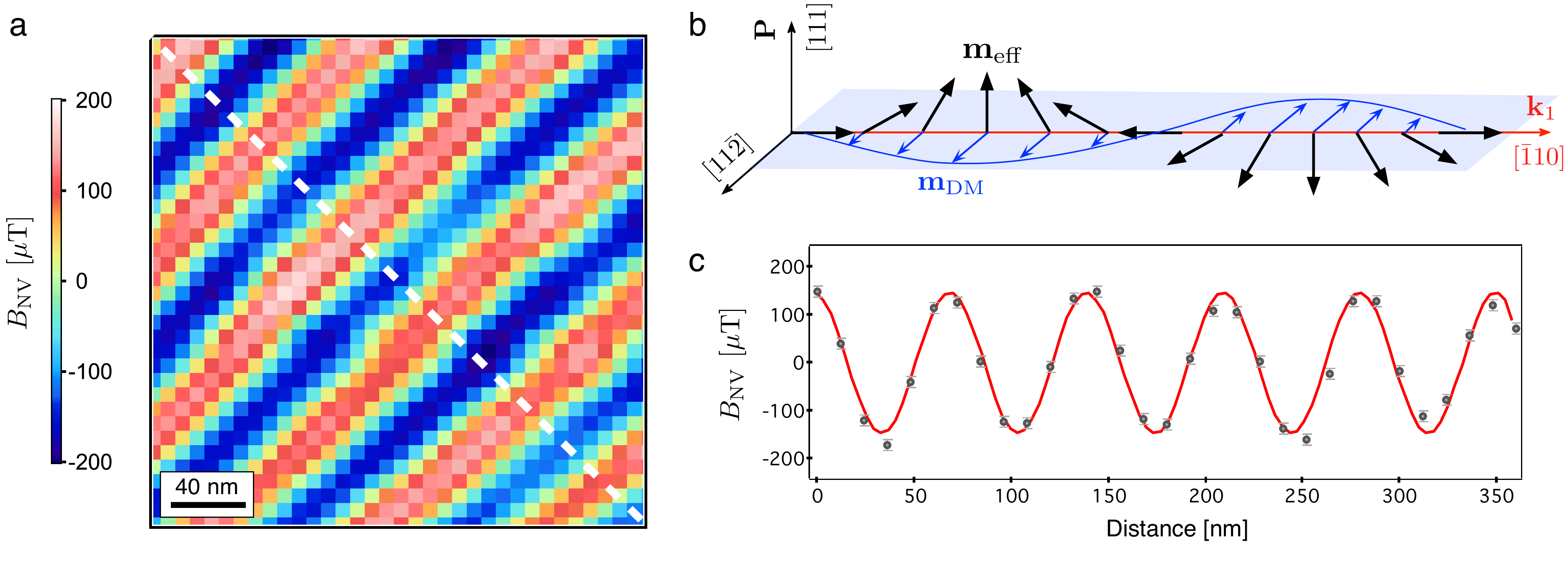}
		\caption{{\bf Figure 4 $|$ Quantitative analysis of the spin cycloid magnetic texture.} {\bf a,} Fully quantitative magnetic field distribution $B_{\rm NV}$ recorded above the ferroelectric monodomain shown in Fig.~3a. {\bf b,} Schematic representation of the spin density wave (SDW) corresponding to an uncompensated magnetic moment $\mathbf{m}_{\rm DM}$ (blue arrows) oscillating in the $[11\bar{2}]$ direction, {\it i.e.} perpendicular to both the ferroelectric polarization vector and $\mathbf{k}_1$. The uncompensated moment due to the pure cycloid $\mathbf{m}_{\rm eff}$ is shown with black arrows. {\bf c,} Linecut of the magnetic field distribution along the cycloid propagation direction (white dashed line in {\bf a}). The black symbols are the experimental data with the standard error (s.e.) while the red solid line is the result of a fit using the analytical formula of the stray field produced by the BFO sample for $d=49$~nm, $m_{\rm eff}=0.07 \ \mu_{\rm B}$, $\lambda=$~$70$~nm, $a=0.396$~nm, and $t=32$~nm. The only free parameter is $m_{\rm DM}$.}
			\end{figure}
			
The Dzyaloshinskii-Moriya (DM) interaction resulting from the alternate rotation of the FeO$_6$ octahedra along the [111] direction is another source of non compensation of the magnetic moments in BFO~\cite{Park,Ederer2005}. In the homogeneous G-type state obtained at high magnetic fields ($>20$~T), this effect is known to generate a weak and uniform magnetization. In the cycloidal state, this magnetization is converted into a spin density wave (SDW) oscillating in the $[11\bar{2}]$ direction, which leads to a periodic wiggling of the cycloidal plane~\cite{Ramazanoglu2011}. As sketched in Fig.~4b, the SDW can be simply modeled by an additional uncompensated magnetization vector $\mathbf{M}_{\rm DM}=\mathbf{m}_{\rm DM}/V$ such that
\begin{equation}
\mathbf{m}_{\rm DM}(\mathbf{r^{\prime}}) = m_{\rm DM}  \cos(\mathbf{k_1}\cdot\mathbf{r^{\prime}}) (\mathbf{e_{k_1}} \times \mathbf{e_P}) \ .
\end{equation}
The value of the SDW amplitude $m_{\rm DM}$ still remains debated. Although it is often considered small ($\sim 0.03 \ \mu_{\rm B}$) or even negligible~\cite{Park}, polarized neutron scattering studies have revealed a maximum amplitude of $0.09 \ \mu_{\rm B}$ in bulk BFO~\cite{Ramazanoglu2011}, which is slightly larger than the uncompensated moment $m_{\rm eff}$ due to the pure cycloid.

An analytical calculation of the stray field produced above the BFO sample is given in Methods. We postulate here that the magnetic structure generating the stray field is a wriggling cycloid as described elsewhere~\cite{Park,Ramazanoglu2011}. The magnetic potential $\Phi$ produced by the magnetization pattern $\mathbf{M}=\mathbf{M}_{\rm DM}+\mathbf{M}_{\rm eff}$ is first calculated using Fourier methods for a monolayer of the BFO sample~\cite{Miranda_PRB2011}. The resulting magnetic field is given by $\mathbf{B}^m=-\nabla \Phi$ and the total  field $\mathbf{B}$ produced at a distance $z$ above the BFO sample surface is finally obtained by summing the contribution from each monolayer. In the laboratory frame $(x,y,z)$ [cf. Extended Data Fig.~5], the stray field components are given by\\
\begin{equation}
\begin{dcases} 
B_x(\mathbf{r})=  -\mathcal{A} \ e^{-k_1 z}\left[C_1 m_{\rm eff}\cos(\mathbf{k_1}\cdot\mathbf{r}) - C_2 m_{\rm DM}\sin(\mathbf{k_1}\cdot\mathbf{r})\right] \\
B_y(\mathbf{r})= \mathcal{A} \ e^{-k_1 z}\left[C_1 m_{\rm eff}\cos(\mathbf{k_1}\cdot\mathbf{r}) - C_2 m_{\rm DM}\sin(\mathbf{k_1}\cdot\mathbf{r})\right] \\
B_z(\mathbf{r})= \sqrt{2}\mathcal{A} \ e^{-k_1 z}\left[C_1 m_{\rm eff}\sin(\mathbf{k_1}\cdot\mathbf{r}) + C_2 m_{\rm DM}\cos(\mathbf{k_1}\cdot\mathbf{r})\right]\ ,
\end{dcases}
\label{eq:B_Bloch} 
\end{equation} 

where $C_1=1+1/\sqrt{3}$, $C_2=2/\sqrt{6}$, and 
\begin{equation}
\mathcal{A}=\frac{\mu_0}{\sqrt{2}V} \ \left[\frac{1-e^{-k_1t}}{1-e^{-k_1a}} \right]\sinh(\frac{ak_1}{2})\ .
\end{equation} 
Here $a$ is the thickness of a BFO monolayer and $t$ the total thickness of the sample. These magnetic field components are then projected along the independently measured NV defect axis in order to obtain an analytical formula for $B_{\rm NV}$. This formula was used to perform a two-dimensional fit of the experimental data while using $m_{\rm DM}$ as the only fitting parameter [Fig.~4c]. A thorough analysis of uncertainties is given in Methods, including those related to (i) the fitting procedure itself, (ii) the probe-to-sample distance $d$, (iii) the cycloid wavelength $\lambda$, (iv) the sample thickness $t$ and (v) the NV defect orientation. This study leads to $m_{\rm DM}=0.16 \pm 0.06 \ \mu_{\rm B}$, where the overall uncertainty of $\sim 40\%$ mainly results from the imperfect knowledge of the probe-to-sample distance [Extended Data Fig. 6c]. We note that the stray field produced above the BFO sample also depends on the chirality of the spin cycloid~\cite{Miranda_PRB2011}. Equation~(\ref{eq:B_Bloch}) is obtained for a spin cycloid with a {\it counter-clockwise} chirality. A similar analysis performed for a {\it clockwise} chirality would lead to a larger amplitude of the SDW, $m_{\rm DM}=0.21 \pm 0.08 \ \mu_{\rm B}$ [see Methods]. In both cases, our study suggests a DM interaction significantly stronger than all reported values in the literature. This result could be explained by considering that the DM interaction is enhanced by the abrupt broken inversion symmetry occurring at the sample surface and then propagated in the BFO thin film by exchange interaction. This observation opens many perspectives for studying emergent interface-induced magnetic interactions resulting from a local breaking of inversion symmetry.

In summary, we have reported the first real-space imaging and electric-field control of the cycloidal antiferromagnetic order in a BFO thin film using a scanning-NV magnetometer operating under ambient conditions. These results open new perspectives for unravelling intriguing phenomena occurring in multiferroic materials like BFO, from magnetoelectric coupling~\cite{Heron}, peculiar properties induced by surface symmetry breaking, to conduction and magnetotransport properties at ferroelectric domain walls~\cite{Balke2012,He_2012}. On a broader perspective, NV magnetometry appears as a unique tool for studying the antiferromagnetic order at the nanoscale. In this way, similar investigations could be extended to a myriad of non-collinear antiferromagnetic materials, or to the domain walls of regular antiferromagnets, opening an exciting avenue towards the development of low-power spintronics~\cite{AFreview}.

\begin{addendum}
 \item [Acknowledgements] We thank J. P. Tetienne and T. Hingant for experimental assistance at the early stage of the project. We are grateful to J. M. D. Coey for fruitful discussions. This research has been supported by the European Research Council (ERC-StG-2014, {\sc Imagine}), the European Union Seventh Framework Program (FP7/2007-2013) under the project {\sc Diadems} and by the French Agence Nationale de la Recherche (ANR) through project {\sc Ferromon}.
 \item[Author contributions] I.G., W.A., L.J.M. and S.C. performed the NV magnetometry experiments; I.G., W.A., L.J.M. and V.J. analyzed the data and performed magnetic modeling with assistance from M.V.; K.G. and C.C. fabricated the BFO sample; V.G. and S.F. performed the structural analysis and the piezoresponse force microscopy experiments; P.A. and P.M. engineered diamond tips hosting single NV defects; I.G., W.A., V.G., S.F., M.B. and V.J. wrote the manuscripts. All authors contributed to the interpretation of the data and commented on the manuscript.
 \item[Competing Interests] The authors declare no competing financial interests.
 \item[Correspondence] Correspondence should be addressed to V. J. (email: vincent.jacques@umontpellier.fr).

\end{addendum}

\noindent {\bf \large Methods}
\vspace{-0.5cm}
\begin{addendum}
 \item [Sample growth.] The epitaxial thin film heterostructure was grown by pulsed laser deposition using a KrF excimer laser ($\lambda = 248$~nm, $1$~J.cm$^{-2}$) on an orthorhombic DyScO$_3$ (110)$_{\rm o}$ single crystal substrate. The SrRuO$_3$ bottom electrode ($1.2$~nm) was grown with $5$~Hz repetition rate at 650$^\circ$C under $0.2$~mbar of oxygen. The BiFeO$_3$ film (32 nm) was subsequently grown at 650$^\circ$C under $0.36$~mbar of oxygen with $1$~Hz repetition rate. The sample was slowly cooled down under high oxygen pressure. The film surface exhibits single-unit-cell atomic steps [Extended Data Fig.~1a]. 
 \item [Structural properties.] We investigated the structural properties of the BiFeO$_3$ thin film by X-ray diffraction (XRD). DyScO$_3$ has an orthorhombic structure$^{31}$ (Pbnm) with $a_{\rm o}=0.5440$~nm, $b_{\rm o}=0.5717$~nm and $c_{\rm o}=0.7903$~nm but can be described in a monoclinic cell on its (110)$_{\rm o}$ orientation$^{32}$. The two in-plane directions are then $\mathbf{a}$ $||$ [001]$_{\rm o}$ and $\mathbf{b}$ $||$ [\={1}10]$_{\rm o}$ and the out-of-plane $\mathbf{c}$ axis is slightly tilted so that $\alpha=2\tan^{-1}(\frac{a_0}{b_0})=87.2^\circ$, $\beta=\gamma =90^\circ$, $a=\frac{c_0}{2} = 0.3952$~nm, $b=c= \frac{\sqrt{a_0^2+b_0^2}}{2} = 0.3947$~nm. In the following, we will only use the monoclinic notation for DyScO$_3$ and BiFeO$_3$. 
 
\indent The $\omega-2\theta$ pattern shows only $(00l)$ peaks for DySc0$_3$ and BiFeO$_3$ indicating that the film is epitaxial and single phase [Extended Data Fig. 1b]. In addition, the presence of Laue fringes  indicates a well-crystallized structure and the peak-to-peak spacing corresponds to a thickness of $32$~nm [Extended Data Fig. 1b]. 

\indent To get more insights into the structure of BiFeO$_3$ thin films, we performed reciprocal space mappings (RSMs) along different directions of the monoclinic DyScO$_3$ substrate [Extended Data Fig.~2]. The films are coherently strained as shown by the same $Q_{x,y}$ as the substrate for $(00l)_{\rm D}$, $(h0l)_{\rm D}$, and $(0kl)_{\rm D}$ RSMs. Furthermore, two Q$_z$ film variants are observed for $(h0l)_{\rm D}$ RSMs and only one for $(0kl)_{\rm D}$ RSMs. Thus, the RSM data are fully consistent with only two monoclinic domains of BiFeO$_3$ epitaxially grown on top of DyScO$_3$ (Ref. 33). For the first one $(001)_{\rm B} \parallel (001)_{\rm D}$ and $[100]_{\rm B} \parallel [110]_{\rm D}$ [green color in Extended Data Fig.~2], while the second one is rotated in plane by 90$^\circ$ so that $(001)_{\rm B} \parallel (001)_{\rm D}$ and $[100]_{\rm B} \parallel [\bar{1}10]_{\rm D}$ [blue color in Extended Data Fig.~2]. We found $\beta=$89.2$^\circ$, $\alpha=\gamma=$90$^\circ$, $a=0.5601$~nm, $b=0.5572$~nm and $c=0.3991$~nm for the structural parameters of BiFeO$_3$ thin films in the monoclinic cell representation. This corresponds to a pseudo-cubic unit-cell volume $V=0.06227$~nm$^3$.

 \item [Ferroelectric properties.] PFM experiments were conducted with an atomic force microscope (Nanoscope V multimode, Bruker) and two external SR830 lock-in detections (Stanford Research) for simultaneous acquisition of in-plane and out-of-plane responses. A DS360 external source (Stanford Research) was used to apply the AC excitation to the SrRuO$_3$ bottom electrode at a frequency of $35$~kHz while the conducting Pt coated tip was grounded. The hysteresis cycle of the out-of-plane PFM is imprinted toward positive bias voltage values [Extended Data Fig. 3a], in accordance with the homogeneous pristine downward polarization detected by out-of-plane PFM imaging [Extended Data Fig. 3b]. 
 
The ferroelectric configuration of the pristine BFO sample was identified by vectorial PFM, {\it i.e.} probing the different in-plane variants when rotating the sample crystallographic axis compared to the PFM cantilever long axis [extended Data Fig.~3c-k]$^{34}$. Alternated light/dark stripes are observed in the in-plane PFM phase image acquired with the cantilever aligned along the pseudo cubic $[100]_{\rm c}$ direction [Extended Data Fig.~3c or Fig.~1d]. This configuration does not lift the degeneracy between equivalent polarization variants for PFM response: all four variants with polarization pointing downwards [sketched in Extended Data Fig. 3e] correspond to the same in plane amplitude response [Extended Data Fig.~3d]. $\mathbf{P}_2^-$ and $\mathbf{P}_3^-$ are pointing to the right of the cantilever, corresponding to the dark phase signal, while $\mathbf{P}_1^-$ and $\mathbf{P}_4^-$ are pointing to the left of the cantilever, corresponding to the light phase signal. At this stage, several kinds of domain walls are still possible (for instance $109^\circ$ domain walls between $\mathbf{P}_1^-$ and $\mathbf{P}_3^-$, or $71^\circ$ domain walls between $\mathbf{P}_1^-$ and $\mathbf{P}_2^-$). When the cantilever is aligned along $[110]_c $ [Extended Data Fig. 3f)], the $\mathbf{P}_2^-$ and $\mathbf{P}_4^-$ in-plane responses are turned off and all responding domains (bright amplitude in Extended Data Fig. 3g) are pointing to the left side of the cantilever (light phase signal in Extended Data Fig. 3h), identifying $\mathbf{P}_3^-$ domains. When the cantilever is aligned along $[\bar{1}10]_c$ (Extended Data Fig. 3i), the $\mathbf{P}_1^-$ and $\mathbf{P}_3^-$ in-plane responses are turned off and all responding domains (bright amplitude in Extended Data Fig. 3j) are pointing to the right of the cantilever (dark phase in Extended Data Fig. 3k), identifying $\mathbf{P}_4^-$ domains. Note that Extended Data Fig.~3g and 3j show complementary responses so that the ferroelectric configuration in BFO thin films that we presented in the manuscript is determined as alternated $\mathbf{P}_3^-$ and $\mathbf{P}_4^-$ variants in the form of stripes separated by $71^\circ$ domain walls.

In written areas [Fig. 3a,d], single ferroelectric domains are reproducibly obtained: the out-of-plane component of the polarization is controlled by the above coercive bias applied between the scanning tip and the bottom SrRuO$_3$ electrode. Moreover, the in-plane component is simultaneously defined thanks to the {\it trailing field} induced by the tip motion along the slow scan axis and aligned along the targeted polarization variant$^{35,36}$.

 \item [Scanning-NV magnetometry.] The experimental setup is described in details in Ref.~37. It combines a tuning-fork-based atomic force microscope (AFM) and a confocal optical microscope (attoAFM/CFM, Attocube Systems), all operating under ambient conditions. The NV spin sensor is located at the apex of a nanopillar in a diamond cantilever which is attached to the AFM head. The procedure for engineering the all-diamond scanning probe tips containing single NV defects used in this work can be found in Ref.~38. Electron spin resonance (ESR) spectroscopy was performed by monitoring the NV defect PL intensity while sweeping the frequency of a RF field generated by a gold stripline antenna directly fabricated onto the BFO sample by e-beam lithography. The NV defect quantization axis was measured by recording the ESR frequency as a function of the amplitude and orientation of a calibrated magnetic field$^{39}$. We obtain spherical angles $(\theta=128\pm1^{\circ},\phi=80\pm1^{\circ})$ in the laboratory frame of reference $(x,y,z)$.
 
Magnetic field images recorded in {\it dual-iso-B} imaging mode are obtained with an integration time of $200$~ms per pixel. The quantitative magnetic field distribution shown in Figure 4a is recorded by measuring the ESR spectrum at each pixel of the scan. This spectrum is composed of 10 bins with a bin size of $2$~MHz and an integration time per bin of $65$~ms, leading to a total acquisition time of $650$~ms per spectrum. The magnetic field image shown in Figure~4a is thus obtained within $\approx$ 20 minutes. Each spectrum is fitted with a Gaussian function in order to infer the Zeeman shift of the ESR frequency, and thus the magnetic field $B_{\rm NV}$. The intrinsic standard error (s.e.) of the magnetic field measurement is in the range of $\sim 10 \ \mu$T [see error bars in Fig.~4c].

 \item [Calibration of the probe-to-sample distance.]
 The distance $d$ between the NV spin sensor and the sample surface was inferred by recording the stray magnetic field produced above the edges of an uniformly magnetized ferromagnetic wire [Extended Data Fig. 4a]. A typical Zeeman-shift profile recorded while scanning the NV defect across the edges of a $500$-nm-wide wire of Pt/Co(0.6nm)/AlO$_x$ is shown in Extended Data Fig. 4b. The probe-to-sample distance $d$ is then extracted by fitting the experimental data following the procedure described in Refs.~(24,40). The result of the fit is indicated as a red solid line in Extended Data Fig. 4b, showing a very good agreement with experimental data. The uncertainty and reproducibility of the fitting procedure was first inferred by fitting a set of independent measurements, leading to a relative uncertainty of $1.5\%$ in probe-to-sample distance. Additional uncertainties induced by those on (i) the NV spin characteristics and (ii) the sample geometry were then carefully analyzed following the method described in Ref.~24, leading to $d=49.0\pm 2.4$~nm. The overall uncertainty is thus on the order of $5\%$.

\item [Stray magnetic field produced by the spin cycloid.]

In this section, we calculate the stray magnetic field produced by the spin cycloid. The general methodology can be summarized as follows. The spin texture of the BFO sample is first modeled by a magnetization vector $\mathbf{M}$ describing a cycloid. The magnetic potential $\Phi$ produced by a single layer of the sample is then computed using Fourier methods and the resulting magnetic field $\mathbf{B}^{m}$ is obtained by using the relation $\mathbf{B}^{m} = -\nabla \Phi$. The total magnetic field  $\mathbf{B}$ produced at a distance $z$ above the sample is then calculated by summing up the contributions from all monolayers. The resulting magnetic field distribution is finally projected along the NV defect axis in order to obtain an analytical formula for $B_{NV}$, which can be used to fit the experimental data. The geometry used for the calculation is schematically depicted in Extended Data Fig. 5.

As introduced in the main text, we consider the uncompensated magnetic moments induced (i) by the pure spin cycloid $\mathbf{m_{\rm eff}}$ and (ii) by the spin density wave $\mathbf{m}_{\rm DM}$ [Fig.~4b]. The resulting spin texture of the BFO sample is modeled by a magnetization vector $\mathbf{M}= (\mathbf{m}_{\rm eff}+\mathbf{m}_{\rm DM})/V$, where 
\begin{eqnarray}
\mathbf{m}_{\rm eff}(\mathbf{r'}) &= &m_{\rm eff} \left[\cos(\mathbf{k_1}\cdot\mathbf{r'})\mathbf{e_{k_1}}  +\sin(\mathbf{k_1}\cdot\mathbf{r'})\mathbf{e_P}\right]\\
\mathbf{m}_{\rm DM}(\mathbf{r'}) & =& m_{\rm DM}  \cos(\mathbf{k_1}\cdot\mathbf{r'}) (\mathbf{e_{k_1}} \times \mathbf{e_P}) \ .
\end{eqnarray}
 A rotation matrix was used to translate this magnetization into the laboratory frame of reference $(x,y,z)$, in which the NV defect quantization axis is defined.

We start by computing the magnetic potential $\Phi(x,y,z)$ produced by a monolayer of the BFO sample, {\it i.e.} with a unit cell thickness $a=0.395$~nm. The magnetic potential is given by$^{29,41}$
\begin{equation}\label{eq:1}
\mathrm{\Phi}(x,y,z) = \iint \limits_{x',y' = -\infty}^{x',y' =+\infty}\, \int \limits_{z' = -a/2}^{z' = +a/2} \, - \frac{\mu_0}{4\pi} \mathbf{M}(x',y') \boldsymbol{\cdot} \nabla (\frac{1}{\sqrt{(x-x')^2 + (y-y')^2 + (z-z')^2}})\, dx' \,dy'\,dz' \ .
\end{equation}

This equation includes a two-dimensional convolution defined as 
\begin{equation}
f(x,y) \ast  g(x,y) = \iint \limits_{x',y' = -\infty}^{x',y' =+\infty}\, f(x',y')\, g(x-x',y-y')\, dx'\,dy' \ ,
\end{equation}
so that the magnetic potential can be expressed as
 \begin{equation}\label{eq:2}
\mathrm{\Phi}(x,y,z)  =  - \frac{\mu_0}{4\pi} \int \limits_{z' = -a/2}^{z' = +a/2} \,\Big[({M_x} \ast \frac{\partial}{\partial x}  \frac{1}{r_0})\,+\, ({M_y} \ast \frac{\partial}{\partial y}  \frac{1}{r_0})\,+\,({M_z} \ast \frac{\partial}{\partial z}  \frac{1}{r_0})\,\Big] dz' \ .
\end{equation}
where 
 \begin{equation}
 \frac{1}{r_0} = \frac{1}{\sqrt{x^2 + y^2 + (z-z')^2}} 
 \end{equation}
 
\noindent and $M_x,M_y,M_z$ are the components of the magnetization. 

Taking the Fourier transform of Eq.~(\ref{eq:2}) and using the convolution theorem $\mathcal{F}[f \ast  g] = \mathcal{F}\,[f]\mathcal{F}[g]$, we obtain
\begin{equation}\label{eq:3}
\mathcal{F}(\mathrm{\Phi})  =  - \frac{\mu_0}{4\pi} \int \limits_{z' = -a/2}^{z' = +a/2} \,\Big[\mathcal{F}({M_x}) \mathcal{F}(\frac{\partial}{\partial x}  \frac{1}{r_0})\,+\, \mathcal{F}({M_y}) \mathcal{F}( \frac{\partial}{\partial y}  \frac{1}{r_0})\,+\,\mathcal{F}({M_z}) \mathcal{F}( \frac{\partial}{\partial z}  \frac{1}{r_0})\,\Big] dz' \ .
\end{equation}
Here $M_x, M_y$ and $M_z$ involve sine and cosine terms whose Fourier transform are given by Dirac $\delta$ function. The Fourier transform of the $(\frac{1}{r})$ terms can be obtained by following the procedure described in Ref.~41. The magnetic potential produced by a monolayer of the sample is finally obtained through an inverse Fourier transform leading to 
 \begin{equation}\label{eq:6}
\mathrm{\Phi}(x,y,z) =  \frac{\mu_0 \sinh(ak_1/2)}{V k_1}   \mathrm{e}^{-k_1 z}\, \Big\{ C_1 m_{\rm eff} \sin(\mathbf{k_1 \cdot r})   +  C_2 m_{\rm DM}   \cos(\mathbf{k_1 \cdot r})    \Big\}
\end{equation}
where $C_1=1+1/\sqrt{3}$ and $C_2=2/\sqrt{6}$. \\

The stray magnetic field $\mathbf{B}^{m}$ produced at a distance $z$ above a monolayer was then calculated using the relation $\mathbf{B}^{m} = -\nabla \Phi$. The resulting stray field components are given by 

\begin{equation}
\begin{dcases} \label{eq:7}
B_x^{m}  =   -\frac{\mu_0 \sinh(ak_1/2)}{\sqrt{2}V}\,  \mathrm{e}^{-k_1 z}\, \Big\{ C_1\, m_{\rm eff}\,  \cos(\mathbf{k_1 \cdot r})   -  C_2\, m_{\rm DM}\,  \sin(\mathbf{k_1 \cdot r})   \Big\}   \\
B_y^{m}  =   +\frac{\mu_0 \sinh(a k_1/2)}{\sqrt{2}V}\,  \mathrm{e}^{-k_1 z}\, \Big\{ C_1\, m_{\rm eff}\,  \cos(\mathbf{k_1 \cdot r})   -  C_2\, m_{\rm DM}\,  \sin(\mathbf{k_1 \cdot r})\Big\} \\
B_z^{m} =    +\frac{\mu_0 \sinh(a k_1 /2)}{V}\, \mathrm{e}^{- k_1z}\, \Big\{ C_1\, m_{\rm eff}\,  \sin(\mathbf{k_1 \cdot r})  +  C_2\, m_{\rm DM}\,  \cos(\mathbf{k_1 \cdot r})\Big\} \ .
\end{dcases}
\end{equation}

The total magnetic field $\mathbf{B}$ produced by the sample is obtained by summing the contribution of each monolayer

\begin{equation}
\begin{dcases}
B_x  = \displaystyle\sum_{j=0}^{N-1}  -\frac{\mu_0 \sinh(a k_1/2)}{\sqrt{2}V}\,  \mathrm{e}^{- k_1 (z + ja)}\, \Big\{  C_1\, m_{\rm eff}\,  \cos(\mathbf{k_1 \cdot r})   -  C_2\, m_{\rm DM}\,  \sin(\mathbf{k_1 \cdot r})  \Big\}   \\
B_y   =  \displaystyle\sum_{j=0}^{N-1} +\frac{\mu_0 \sinh(a k_1/2)}{\sqrt{2}V}\,  \mathrm{e}^{-k_1 (z + ja)}\, \Big\{    C_1\, m_{\rm eff}\,  \cos(\mathbf{k_1 \cdot r})   -  C_2\, m_{\rm DM}\,  \sin(\mathbf{k_1 \cdot r})  \Big\} \\
B_z   =   \displaystyle\sum_{j=0}^{N-1} +\frac{\mu_0 \sinh(a k_1/2)}{V}\,  \mathrm{e}^{- k_1 (z + ja)}\, \Big\{   C_1\, m_{\rm eff}\,  \sin(\mathbf{k_1 \cdot r})  +  C_2\, m_{\rm DM}\,  \cos(\mathbf{k_1 \cdot r})   \Big\} \ ,
\end{dcases}
\end{equation}
where $N$ is the number of atomic layer of the BFO sample. 

The above equation can be further simplified as
\begin{equation}
\begin{dcases}
B_x =    - \mathcal{A}\, \mathrm{e}^{-k_1 z}\, \Big\{C_1\, m_{\rm eff}\, \cos(\mathbf{k_1 \cdot r})    -  C_2\, m_{\rm DM}\, \sin(\mathbf{k_1 \cdot r})  \Big\}   \\
B_y=    +\mathcal{A}\, \mathrm{e}^{-k_1z}\, \Big\{  C_1\, m_{\rm eff}\, \cos(\mathbf{k_1 \cdot r})  -  C_2\, m_{\rm DM}\, \sin(\mathbf{k_1 \cdot r}) \Big\}   \\
B_z=    + \sqrt{2} \mathcal{A}\, \mathrm{e}^{-k_1 z}\, \Big\{ C_1\, m_{\rm eff} \,\sin(\mathbf{k_1 \cdot r})    +  C_2\, m_{\rm DM}\, \cos(\mathbf{k_1 \cdot r})   \Big\}   
\end{dcases}
\label{ccw}
\end{equation}
where  
\begin{equation}
\mathcal{A}=\frac{\mu_0}{\sqrt{2}V} \ \left[\frac{1-e^{-k_1t}}{1-e^{-k_1a}} \right]\sinh(\frac{ak_1}{2})\ .
\end{equation} 
Here $t=Na$ is the total thickness of the BFO sample. 

This magnetic field distribution is finally projected along the NV defect axis in order to obtain an analytical formula for $B_{\rm NV}$, which is given 
\begin{equation}
B_{\rm NV}=B_x\cos\phi\sin\theta + B_y\sin\phi\sin\theta + B_z\cos\theta \ ,
\label{Final}
\end{equation} 
where ($\theta,\phi$) are the spherical angles of the NV axis in the laboratory frame. 

Equation~(\ref{Final}) was used to perform a two-dimensional fit of the experimental data, while using $m_{\rm DM}$ as the {\it only free parameter}. The quality of the fit is illustrated by Extended Data Fig.~6b.

\item [Analysis of uncertainties.]

We now analyze the uncertainty on the fit outcome $m_{\rm DM}$, which results (i) from the fitting procedure and (ii) from uncertainties on the parameters $p_i=\{\lambda,m_{\rm eff},t,d,\theta,\phi\}$ that are involved in the expression of $B_{\rm NV}$. In the following, the parameters $p_i$ are expressed as $p_i=\bar{p_i}+\sigma_{p_i}$ where $\bar{p_i}$ denotes the nominal value of parameter $p_i$ and $\sigma_{p_i}$ its standard error. These parameters, which are summarized in Extended Data Fig.~6c, are evaluated as follows :
\begin{itemize}
\item[$\bullet$] \ The cycloid wavelength $\lambda$ can be precisely extracted through an independent two-dimensional fit of the experimental data with a simple sinusoidal function. We obtain $\lambda=70.0\pm1.4$~nm for the quantitative magnetic field image shown in Figure 4. The uncertainty ($\sim 2\%$) comes from the calibration of the $(x,y)$ scanner.

\item[$\bullet$] \ From the measured cycloid wavelength, we infer a characteristic canting angle of $\frac{360{^\circ}}{\lambda}=$~$5.14 \ \pm$~$0.10 {^\circ}$/nm, leading to $\alpha_c=2.04\pm0.02^{\circ}$ between neighboring antiferromagnetically coupled Fe atoms, which are separated by $a=0.395$~nm [see Fig.~1b]. The resulting uncompensated magnetic moment per Fe atom is given by $m_{\rm eff}=m_{\rm Fe}\sin(\alpha_c/2)=0.073\pm 0.001 \  \mu_{\rm B}$. 

\item[$\bullet$] \ The thickness of the BFO sample is extracted through X-ray diffraction (XRD) measurements [see Extended Data Fig.~1b]. The peak-to-peak spacing of Laue fringes indicates a sample thickness $t=32\pm2$~nm.

\item[$\bullet$] \ The NV defect quantization axis is measured by recording the ESR frequency as a function of the amplitude and orientation of a calibrated magnetic field, leading to spherical angles $(\theta=128\pm1^{\circ},\phi=80\pm1^{\circ})$ in the laboratory frame $(x,y,z)$. 

\item[$\bullet$] \ The probe-to-sample distance $d$ is inferred through a calibration measurement described in the previous section, leading to $d=49.0\pm 2.4$~nm.
\end{itemize}

We first evaluate the uncertainty of the fitting procedure. To this end, a two-dimensional fit of the experimental data was performed with Equation~(\ref{Final}) while fixing all the parameters $p_i$ to their nominal values $\bar{p_{i}}$, leading to $m_{\rm DM}=0.160\pm0.002 \ \mu_{\rm B}$. The relative uncertainty linked to the fitting procedure is therefore given by $\epsilon_{\rm fit}=1.2 \%$. We note that the intrinsic accuracy of the magnetic field measurement is in the range of $\delta B_{\rm NV}\sim$~$10 \ \mu$T. This leads to an uncertainty of the SDW amplitude $\delta m_{\rm DM}\sim 0.01 \ \mu_{\rm B}$, corresponding to a relative uncertainty $\epsilon_{\rm m}=6 \%$.  

In order to estimate the relative uncertainty $\epsilon_{p_{i}}$ introduced by each parameter $p_i$, the two-dimensional fit was performed with one parameter $p_i$ fixed at $p_i=\bar{p_{i}} \pm \sigma_{p_{i}}$, all the other five parameters remaining fixed at their nominal values. The corresponding fit outcomes are denoted $m_{\rm DM}(\bar{p_{i}} + \sigma_{p_{i}})$ and $m_{\rm DM}(\bar{p_{i}} - \sigma_{p_{i}})$. The relative uncertainty introduced by the errors on parameter $p_i$ is then finally defined as 
\begin{equation} \label{partial_uncert}
\epsilon_{p_{i}}=\frac{m_{\rm DM}(\bar{p_{i}} + \sigma_{p_{i}})-m_{\rm DM}(\bar{p_{i}} - \sigma_{p_{i}})}{2m_{\rm DM}(\bar{p_{i}} )} \ .
\end{equation}
This analysis was performed for each parameter $p_i$ and the resulting uncertainties $\epsilon_{p_{i}}$ are summarized in Extended Data Fig.~6c. The cumulative uncertainty $\epsilon$ is finally given by 
\begin{equation}
\epsilon=\sqrt{\epsilon^{2}_{\rm fit}+\epsilon^{2}_{\rm m}+\sum_i \epsilon_{p_i}^2} \ ,
\label{EqUncert}
\end{equation}
where all errors are assumed to be independent. 

We finally obtain $\epsilon=41\%$ and $m_{\rm DM}=0.16\pm 0.06 \ \mu_{\rm B}$. We note that the dominating source of uncertainty is given by the imperfect knowledge of the probe-to-sample distance ($\epsilon_{d}=39\%$).

\item [Comparison with numerical simulations.]
The only assumption used for the calculation of the stray field above the BFO sample consists in considering a two-dimensional spin texture with infinite size in the ($x^{\prime},y^{\prime}$) plane [cf. Extended Data Fig.~5]. Such an assumption is valid since the typical dimension of the ferroelectric monodomain is in the range of $1\ \mu$m, which is much larger than the probe-to-sample distance $d$. This was further verified by comparing the result of the calculation with numerical simulations. To this end, the magnetization $\mathbf{M}$ of the BFO sample was discretized into uniformly magnetized computation cells with a characteristic mesh volume $1\times 1\times1$~nm$^3$. The magnetic field distribution produced by each magnetization cell was computed at a distance $d$ above the sample surface using standard magnetostatic theory$^{37}$. By summing the contributions of all cells and then projecting along the NV defect quantization axis, we finally obtain a simulation of the stray field distribution $B_{\rm NV}$. Such a numerical simulation is in excellent agreement with the analytical calculation with a deviation smaller than 1$\%$.

\item [Effect of the cycloid chirality.]

In the previous sections, the calculation of the stray field above the BFO sample was performed for a spin cycloid with a counterclockwise (c-cw) chirality. It was emphasized in Ref.~29 that the stray field depends on the chirality of the spin cycloid. By considering a clockwise (cw) chirality, the uncompensated magnetic moment induced by the spin cycloid is modified as 
\begin{equation}
\mathbf{m}_{\rm eff}^{\rm (cw)}(\mathbf{r'}) = m_{\rm eff} \left[\cos(\mathbf{k_1}\cdot\mathbf{r'})\mathbf{e_{k_1}}  -\sin(\mathbf{k_1}\cdot\mathbf{r'})\mathbf{e_P}\right]
\end{equation}

On the other hand, the magnetization distribution resulting from the SDW is independent of the chirality. The resulting magnetic field distribution is then given by
\begin{equation}
\begin{dcases}
B_x =    - \mathcal{A}\, \mathrm{e}^{-k_1 z}\, \Big\{C_1^{\rm (cw)}\, m_{\rm eff}\, \cos(\mathbf{k_1 \cdot r})    -  C_2\, m_{\rm DM}\, \sin(\mathbf{k_1 \cdot r})  \Big\}   \\
B_y=    +\mathcal{A}\, \mathrm{e}^{-k_1z}\, \Big\{  C_1^{\rm (cw)}\, m_{\rm eff}\, \cos(\mathbf{k_1 \cdot r})  -  C_2\, m_{\rm DM}\, \sin(\mathbf{k_1 \cdot r}) \Big\}   \\
B_z=    + \sqrt{2} \mathcal{A}\, \mathrm{e}^{-k_1 z}\, \Big\{ C_1^{\rm (cw)}\, m_{\rm eff} \,\sin(\mathbf{k_1 \cdot r})    +  C_2\, m_{\rm DM}\, \cos(\mathbf{k_1 \cdot r})   \Big\}   
\label{cw}
\end{dcases}
\end{equation}
where $C_1^{\rm (cw)}=1-1/\sqrt{3}$. The only difference between this magnetic field distribution and the one obtained for a counterclockwise chirality is the constant $C_1^{\rm (cw)}$ [see Eq.~(\ref{ccw})]. Since $C_1^{\rm (cw)}<C_1$, the stray magnetic field produced by the pure cycloid is weaker for the clockwise chirality. Fitting the experimental data with such a chirality of the spin cycloid leads to $m_{\rm DM}= 0.21\pm 0.08 \ \mu_{\rm B}$. The cycloid chirality could be measured in future experiments by analyzing the stray field amplitude on each side of a single ferroelectric domain wall. In this work, postulating either chirality leads to the similar conclusion that the DM interaction is significantly stronger than all reported values in the literature.

\item [Data availability.] The data that support the findings of this study are available from the corresponding author upon reasonable request.

\end{addendum}
  
{\small
\noindent 31. Uecker, R. {\it et al.} Properties of rare-earth scandate single crystals (Re=Nd-Dy). {\it J. Cryst. Growth} {\bf 310}, 2649-2658 (2008).

\noindent 32. Johann, F., Morelli, A., Biggemann, D., Arredondo, M. \& Vrejoiu, I. Epitaxial strain and electric boundary condition effects on the structural and ferroelectric properties of BiFeO$_3$ films. {\it Phys. Rev. B} {\bf 84}, 94105 (2011).

\noindent 33. Chen, Z. H., Damodaran, A. R., Xu, R., Lee, S. \& Martin, L. W. Effect of Ôsymmetry mismatchÕ on the domain structure of rhombohedral BiFeO3 thin films. {\it Appl. Phys. Lett.} {\bf 104}, 2012-2016 (2014).

\noindent 34. Zavaliche, F. {\it et al.} Multiferroic BiFeO$_3$ films: domain structure and polarization dynamics. {\it Phase Transitions A Multinatl. J.} {\bf 79}, 991-1017 (2006).

\noindent 35. Balke, N. {\it et al.} Deterministic control of ferroelastic switching in multiferroic materials. {\it Nature Nanotech.} {\bf 4}, 868-875 (2009).

\noindent 36. Crassous, A., Sluka, T., Tagantsev, A. K. \& Setter, N. Polarization charge as a reconfigurable quasi-dopant in ferroelectric thin films. {\it Nature Nanotech.} {\bf 10}, 614-618 (2015).
  
\noindent 37. Rondin, L. {\it et al.} Nanoscale magnetic field mapping with a single spin scanning probe magnetometer. {\it Appl. Phys. Lett.} {\bf 100}, 153118 (2012).

\noindent 38. Appel, P., Neu, E., Ganzhorn, M., Barfuss, A., Batzer, M., Gratz, M., Tschoepe, A. \& Maletinsky, P. Fabrication of all diamond scanning probes for nanoscale magnetometry. {\it Rev. Sci. Instrum.} {\bf 87}, 063703 (2016).

\noindent 39. Rondin, L. {\it et al.} Stray-field imaging of magnetic vortices with a single diamond spin. {\it Nature Commun.} {\bf 4}, 2279 (2013).

\noindent 40. Hingant, T., Tetienne, J.-P., Mart\'{\i}nez, L. J., Garcia, K., Ravelosona, D., Roch, J.-F. \& Jacques, V. Measuring the Magnetic Moment Density in Patterned Ultrathin Ferromagnets with Submicrometer Resolution. {\it Phys. Rev. Applied} {\bf 4}, 014003 (2015).

\noindent 41. Blakely, R. J. Potential Theory in Gravity and Magnetic Applications. (Cambridge University Press, 1996).

}

\begin{figure}
		\centering
		\includegraphics[scale=0.5]{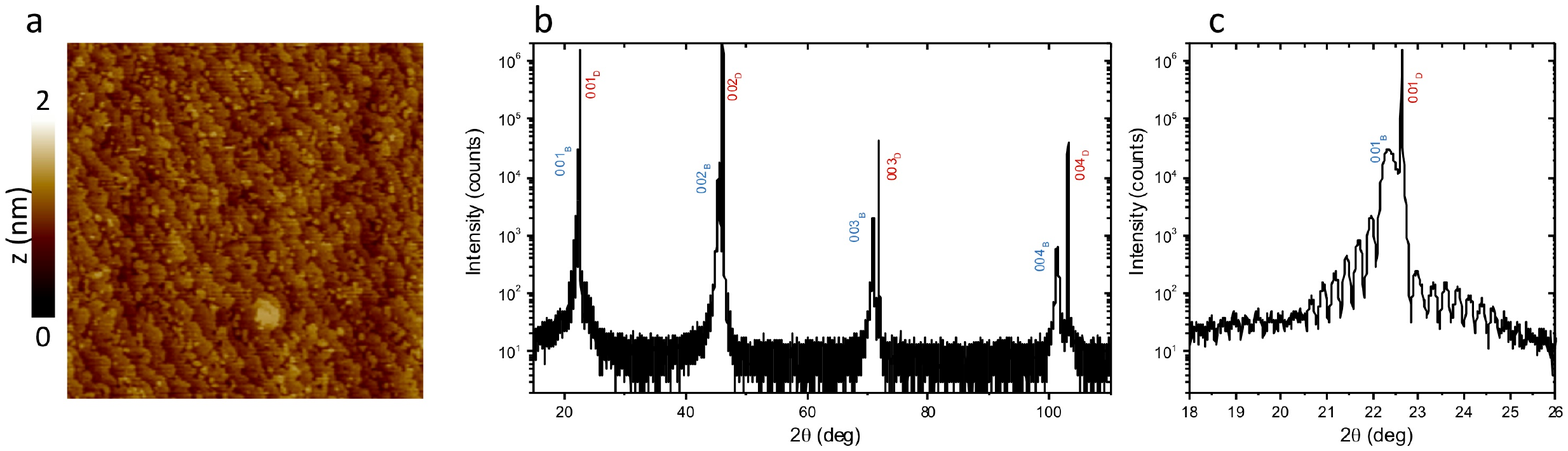}
		{\bf Extended Data Figure 1 $|$ Structural properties. a,} $6\times6 \ \mu$m$^2$ image of the surface topography of the $32$-nm-thick BiFeO$_3$ thin film grown on DyScO$_3$ substrate showing single-unit-cell atomic steps. {\bf b,} X-ray diffraction $\omega-2\theta$ pattern of the same film displays only $(00l)$ peaks for BiFeO$_3$ and DyScO$_3$ (in monoclinic notation). {\bf c,} Zoom along the (001) peak of DyScO$_3$ showing clear Laue fringes.
\end{figure}

\begin{figure}
		\centering
		\includegraphics[scale=0.45]{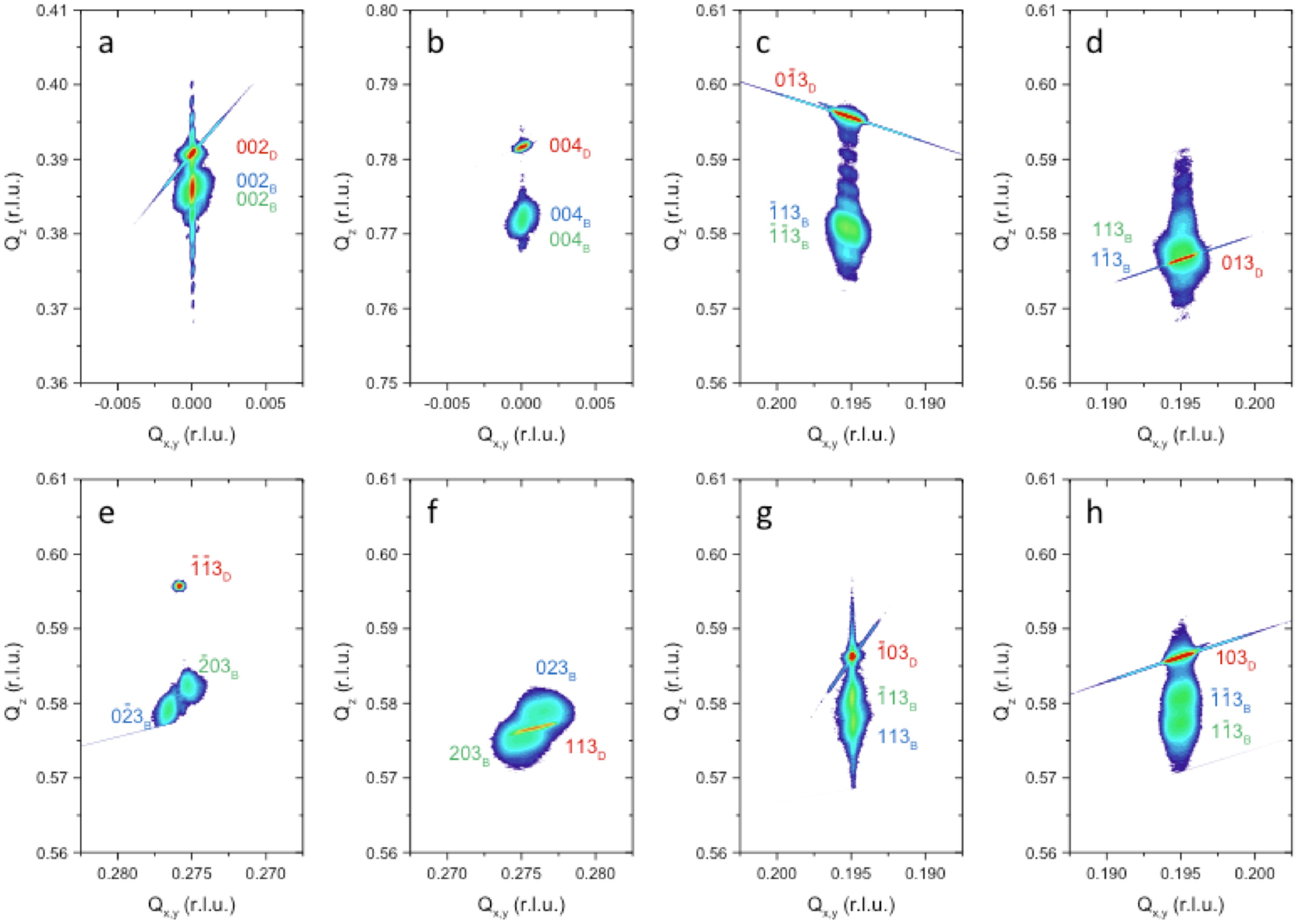}\\
	{\bf Extended Data Figure 2 $|$ Reciprocal space mappings (RSMs) of the  32-nm-thick BiFeO$_3$ film grown on SrRuO$_3$/DyScO$_3$}. RSMs around {\bf a,} $(002)_{\rm D}$, {\bf b,} $(004)_{\rm D}$, {\bf c,} $(0\bar{1}3)_{\rm D}$, {\bf d,} $(013)_{\rm D}$, {\bf e,} $(\bar{1}\bar{1}3)_{\rm D}$, {\bf f,} $(113)_{\rm D}$, {\bf g,} $(1\bar{0}3)_{\rm D}$ and {\bf h,} $(103)_{\rm D}$ planes of DyScO$_3$. All the planes are indexed in monoclinic notation and the subscripts D and B correspond to DyScO$_3$ and BiFeO$_3$, respectively. Two different domains can be identified for monoclinic BiFeO$_3$ (green and blue).
\end{figure}

\begin{figure}
		\centering
		\includegraphics[scale=0.5]{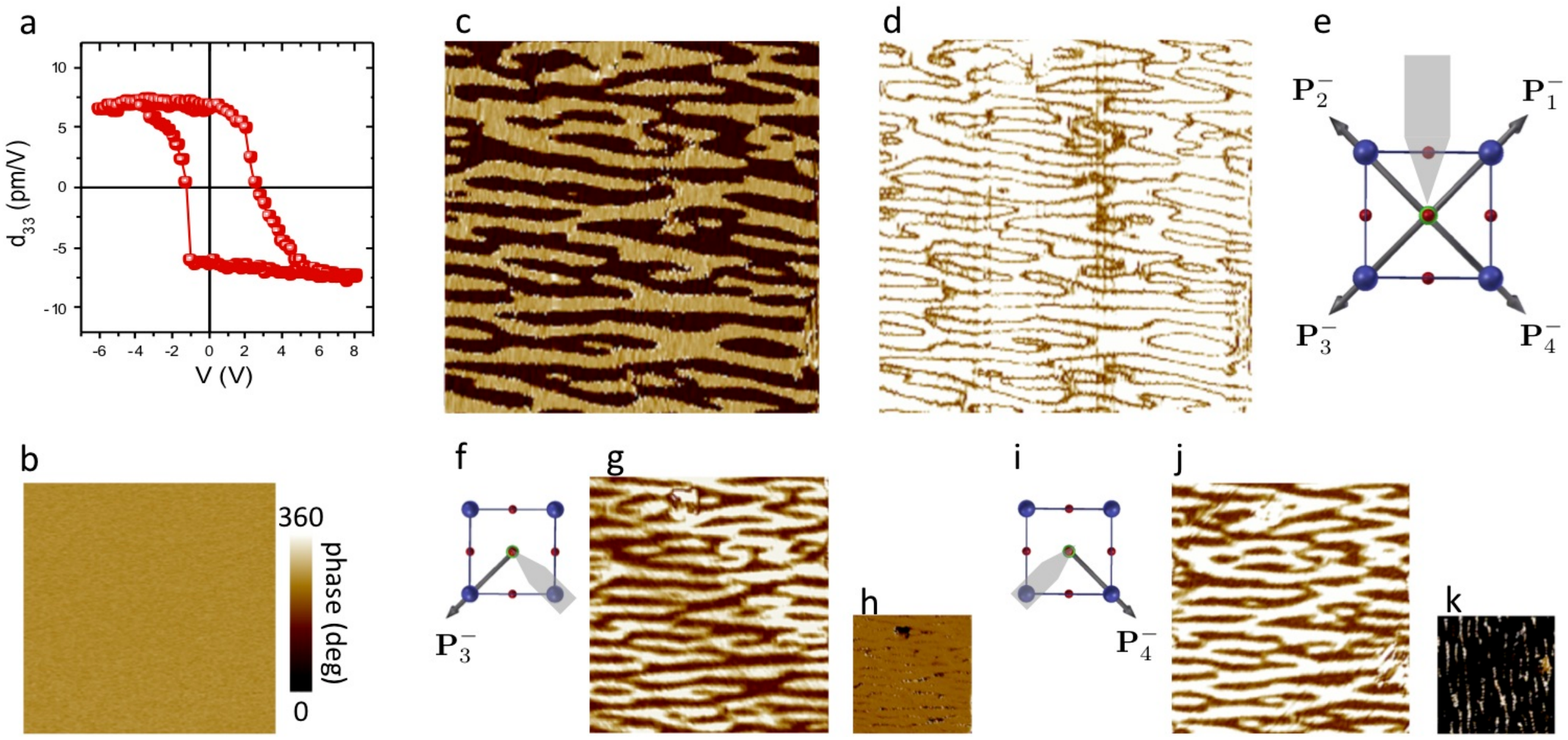}
		{\bf Extended Data Figure 3 $|$ Determination of polarization variants in BFO thin films.} {\bf a,} Local out-of-plane PFM hysteresis loop with bias voltage. {\bf b,} Homogeneous out-of-plane PFM phase corresponding to polarization variants pointing downward in a $6\times6 \ \mu$m$^2$ area. {\bf c,} In-plane PFM phase and {\bf d,} amplitude for the cantilever parallel to $[100]_{\rm c}$. {\bf e,} Sketch of the PFM cantilever and the four possible in plane variants of polarization in BFO. {\bf f,} Sketch of the $[110]_{\rm c}$ direction of the cantilever with the corresponding in-plane PFM {\bf g,} amplitude and {\bf h,} phase. {\bf i,} Sketch of the $[\bar{1}10]_{\rm c}$ direction of the cantilever with the corresponding in-plane PFM {\bf j,} amplitude and {\bf k,} phase. All the images in {\bf c} to {\bf k} are acquired in the same $3\times3 \ \mu$m$^2$ area.
\end{figure}

\begin{figure}
		\centering
		\includegraphics[scale=0.45]{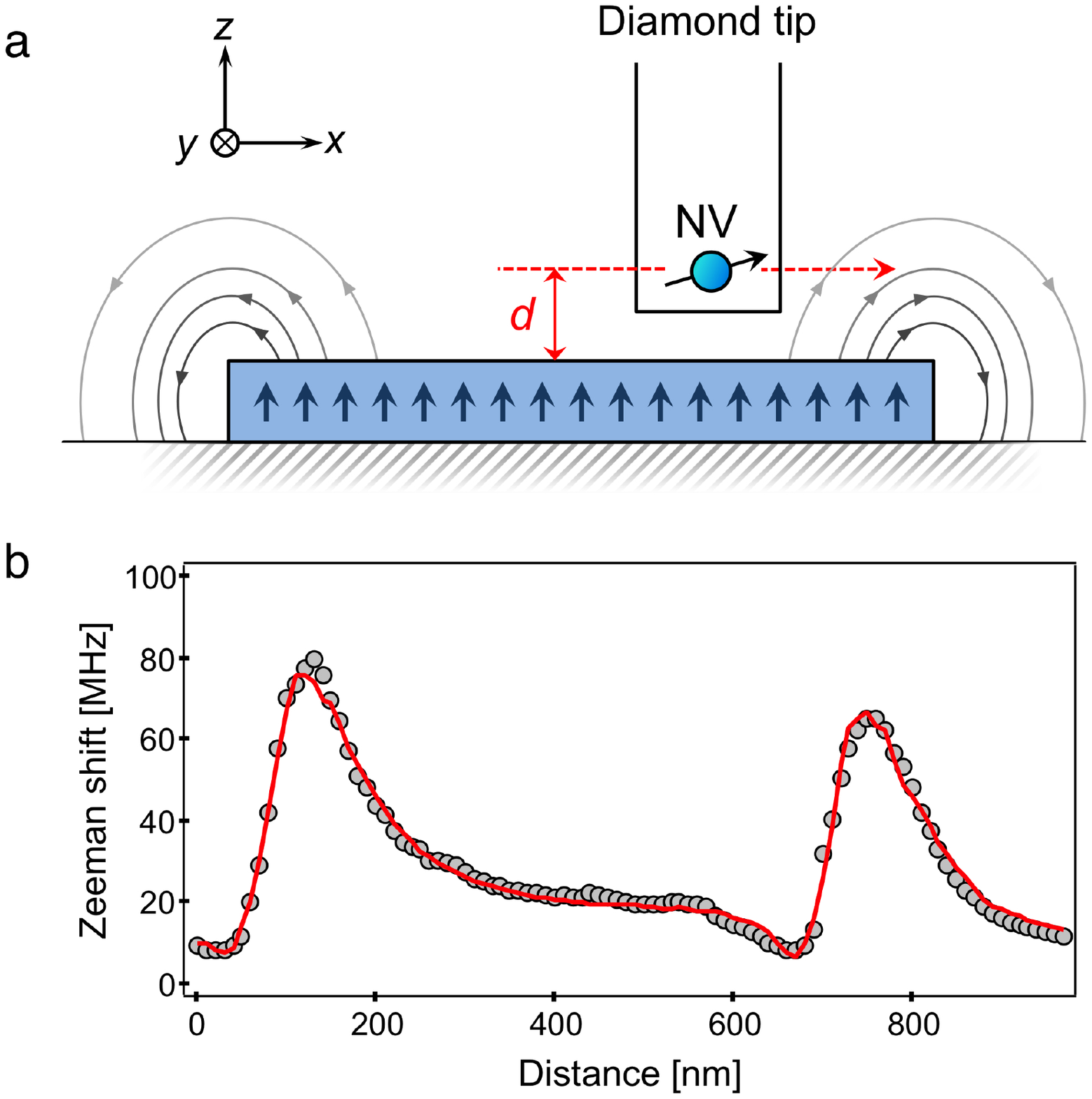}\\
	{\bf Extended Data Figure 4 $|$ Measurement of the probe-to-sample distance}. {\bf a,} The scanning-NV magnetometer is used to measure the magnetic field (grey arrows) produced at the edges of an uniformly magnetized ferromagnetic wire (blue arrows).  {\bf b,} Typical Zeeman-shift profile measured by scanning the NV defect across the edges of a 500-nm-wide wire of Pt/Co(0.6nm)/AlO$_x$ with perpendicular magnetic anisotropy. The markers are experimental data and the red solid line is data fitting from which $d$ is extracted$^{24,40}$. We note that only the absolute value of the magnetic field is measured in this experiment.
\end{figure}

\begin{figure}
		\centering
		\includegraphics[scale=0.35]{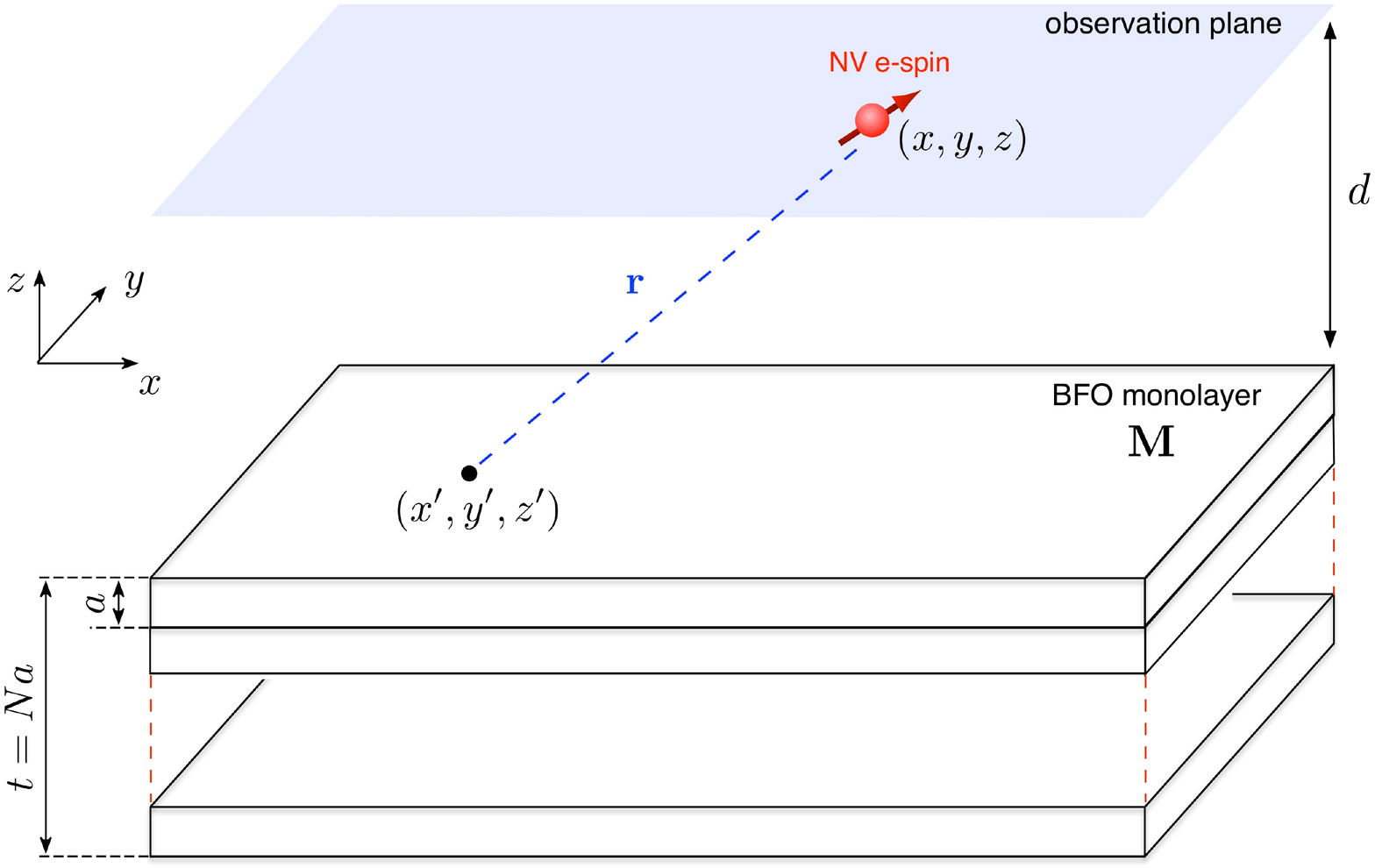}\\
	{\bf Extended Data Figure 5 $|$ Schematic of the geometry used for the stray field calculation.}
\end{figure}

\begin{figure}
		\centering
		\includegraphics[scale=0.35]{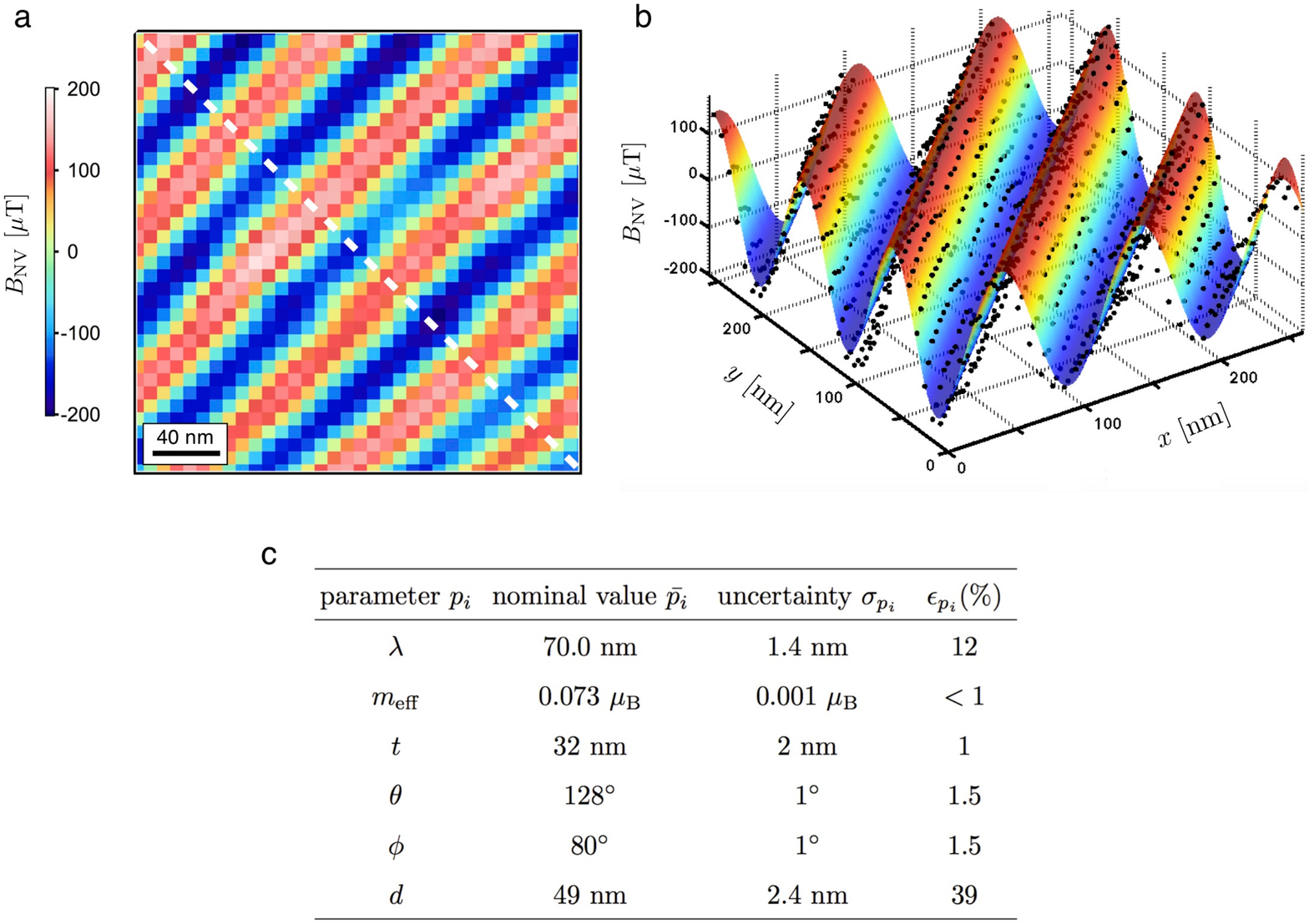}\\
		{\bf Extended Data Figure 6 $|$ Data fitting and uncertainty analysis.} {\bf a,} Magnetic field distribution reproduced from Figure~4a of the main text. {\bf b,} The blacks symbols are the experimental data and the colored solid curve is the result of a two-dimensional fit using Equation~(\ref{Final}) with $d=49$~nm, $m_{\rm eff}=0.07 \ \mu_{\rm B}$, $\lambda=$~$70$~nm, $a=0.396$~nm, $t=32$~nm and $(\theta,\phi)=(128^{\circ},80^{\circ})$. The linecut shown in Figure~4c of the main text correspond to the white dashed line in {\bf a}. {\bf c,} Summary of the relative uncertainties $\epsilon_{p_i}$ on the fitting parameter $m_{\rm DM}$ for the six parameters $p_i=\{\lambda,m_{\rm eff},t,d,\theta,\phi\}$ [cf. Methods].
\end{figure}

\end{document}